# Ultrafast Energy Transfer Between Molecular Assemblies and Surface Plasmons in the Strong Coupling Regime

Maxim Sukharev[1], Tamar Seideman[2], Robert J. Gordon[3], Adi Salomon[4,5], and Yehiam Prior[5]

[1]*School of Letters and Sciences, Arizona State University, Mesa AZ 85212*
*e-mail address: maxim.sukharev@asu.edu*
[2]*Department of Chemistry, Northwestern University, 2145 Sheridan Rd., Evanston IL 60201*
[3]*Department of Chemistry, University of Illinois at Chicago, 845 W. Taylor St., Chicago IL 60680*
[4]*Department of Chemistry, Bar-Ilan University, Ramat-Gan, 52900, Israel*
[5]*Department of Chemical Physics, Weizmann Institute of Science, 76100 Rehovot, Israel*

The nonlinear optical dynamics of nano-materials comprised of plasmons interacting with quantum emitters is investigated by a self-consistent model based on the coupled Maxwell-Liouville-von Neumann equations. It is shown that ultra-short resonant laser pulses significantly modify the optical properties of such hybrid systems. It is further demonstrated that the energy transfer between interacting molecules and plasmons occurs on a femtosecond time scale and can be controlled with both material and laser parameters.

## I. INTRODUCTION

Noble metals, especially nano-structures, are well known for their unique optical properties stemming from the phenomenon of surface plasmon-polariton (SPP) resonances[1]. This research area, known as nanoplasmonics, has grown rapidly in recent years, mainly because of potential applications of plasmonic materials[2]. Extreme concentration of electromagnetic (EM) radiation in nanoscale spatial regions was proposed[3] and implemented experimentally[4] as a method to achieve lasing. Other notable realizations of light EM field localization include surface-enhanced Raman spectroscopy[5] and tip-enhanced resonance microscopy[6].

Among many exciting developments of nanoplasmonics lies the newly emerging research field of nanoscale optical molecular physics, which deals with ensembles of quantum particles optically coupled to nano-materials[7] such as metal nanoparticles[8] (NP) and one- or two-dimensional periodic plasmonic arrays[9]. It has been shown, both theoretically[10] and experimentally[11], that proper utilization of the optical properties of the metal nanostructures may lead to single atom/molecule optical trapping[10a] as well as alignment and focusing. It is now possible[12] to control the geometry of nano-materials (e.g., NP shape, dimensions, and relative arrangement) with a precision on the order of 1 nm. This fine spatial control presents the key for successful manipulation of individual atoms and molecules[10b]. The basis for this manipulation lies in the strong environmental[13] spatial dependence of the evanescent EM field, which generates large field gradients suitable for optical trapping, focusing, and alignment[14]. For example, it has been long known that local EM fields associated with metal NP dimers depend significantly on particle sizes and particle-to-particle distances[15].

Despite considerable progress, our understanding of the optics of quantum media coupled to nano-materials is still incomplete. Many recent works consider few quantum emitters driven by localized EM near-fields in plasmonic materials[16], with only limited attempts to include collective effects[17], which may play a critical role in the quantum optics of nano-materials. Moreover, the optics of hybrid materials comprised of resonant microcavities and ensembles of quantum emitters (quantum dots[18] (QD), molecular ag-gregates[19], nanocrystals[20], and other dipoles[20a, 21]) have been a subject of extensive research in the past several years[22]. For example, it has been demonstrated[23] that the transmission and reflection spectra of a gold film are significantly modified by the deposition of a layer of J-aggregates on the film's surface. It was also shown experimentally[19c] that SPP resonances have a large effect on the molecular electronic structure, leading to Rabi splitting of resonance peaks[24]. This phenomenon was proposed as a way of controlling the optics of hybrid materials with femtosecond laser pulses[25]. Furthermore, core-shell metal NPs with a shell comprised of optically active molecules have been recently studied experimentally[26], demonstrating optimization of the coupling between J-aggregates and a localized SPP, which resulted in Rabi splitting as large as 200 meV. While experimental studies have clearly demonstrated the importance and unique optical properties of hybrid materials[27], there remains a notable gap between the experimental progress and the status of theory and modeling of such systems.

The major parameters determining the strength of the interaction between molecular assemblies and SPP waves are the molecular concentration, transition dipole moment, and local field distribution enhanced by the nano-structure. The strong coupling regime may be defined as cases where the field-induced Rabi splitting of the hybrid system surpasses all linewidths caused by the various damping rates[28]. Strong coupling manifests itself as an avoided crossing of the polariton modes when the plasmon frequency is varied, with a pronounced Rabi splitting that is a non-negligible fraction of the molecular transition frequency[19c, 23-24, 29]. In this regime, energy exchange between the molecular and SPP modes is observed, giving rise to two new polariton eigenmodes. These states have mixed SPP-molecular properties that could be explored and utilized in various applications[26]. Most of the modeling of such systems was done with EM fields under SPP resonant conditions being used as an input for determining subsequent quantum dynamics of a molecular subsystem[28]. At high molecular concentrations, however, this approach is no longer valid because it fails to account for collective effects (e. g., back action of the molecular dipole radiation on the local EM field, which in turn influences the molecules). Furthermore, it was shown recently that proper self-consistent modeling could explain the presence of an additional mode with mixed molecular-plasmon characteristics appearing in the transmission spectra of hybrid materials[30].

In the current work we utilize a self-consistent model based on the Maxwell-Liouville-von Neumann equations and examine the nonlinear dynamics of a hybrid nano-material comprised of a molecular layer optically coupled to a periodic array of sliver slits. We propose a simple method to simulate transient spectroscopy data for hybrid systems and show that in the strong coupling regime energy transfer occurs at the femtosecond time scale. Moreover we demonstrate that the energy distribution can be controlled via laser and material parameters.

The paper is organized as follows. Section II outlines the model and discusses the numerical approach used in our simulations. Section III considers simulations of the linear and nonlinear optical properties of a thin molecular layer. In Sections IV and V we present transient spectroscopy calculations for a system of molecules interacting with a periodic slit array. Section VI summarizes the results and provides an outlook for future research directions.

## II. MODEL AND NUMERICAL IMPLEMENTATION

The interaction of EM radiation with molecular ensembles is treated using a semi-classical model based on the Maxwell-Liouville-von Neumann equations. The dynamics of the EM field, $\vec{H}$ and $\vec{E}$, is governed by the classical Maxwell's equations,

$$\mu_0 \frac{\partial \vec{H}}{\partial t} = -\nabla \times \vec{E},$$
$$\varepsilon_0 \frac{\partial \vec{E}}{\partial t} = \nabla \times \vec{H} - \frac{\partial \vec{P}}{\partial t}, \qquad (1)$$

where $\mu_0$ and $\varepsilon_0$ are the magnetic permeability and dielectric permittivity of free space, respectively. The macroscopic polarization $\vec{P}$ is calculated according to

$$\vec{P} = n_0 \langle \vec{d} \rangle = n_0 \operatorname{Tr}(\hat{\rho} \vec{d}), \qquad (2)$$

where the density matrix $\hat{\rho}$ is the solution of the Liouville-von Neumann equation written in the Lindblad form,

$$\frac{d\hat{\rho}}{dt} = -\left(i/\hbar\right)\left[\hat{H},\hat{\rho}\right] + \sum_n \frac{\gamma_n}{2}\left(2\hat{\sigma}_n^{(-)}\hat{\rho}\hat{\sigma}_n^{(+)} - \hat{\sigma}_n^{(+)}\hat{\sigma}_n^{(-)}\hat{\rho} - \hat{\rho}\hat{\sigma}_n^{(+)}\hat{\sigma}_n^{(-)}\right). \qquad (3)$$

In Eq. (3), $-\left(i/\hbar\right)\left[\hat{H},\hat{\rho}\right]$ is the unitary part of the quantum evolution, $\hat{H}$ being the complete Hamiltonian, $\hat{\sigma}_n^{-/+}$ are the lowering and raising operators, and $\gamma_n$ is the rate at which state $|n\rangle$ decays to the ground state $|1\rangle$ (typically referred to as a $T_1$ process). The dephasing rates (typically referred to as $T_2$ processes) are included in (3) in the off-diagonal terms. The relaxation processes are considered to be Markovian.

To include the back action of the molecules on the field, which is necessary, for example, at high molecular density we introduce the Lorentz-Lorenz correction term for the local electric field in the form[31]

$$\vec{E}_{\text{local}} = \vec{E} + \frac{1}{3\varepsilon_0}\vec{P}. \qquad (4)$$

It had been shown[32] that this local field correction (4) is valid also in case of EM wave propagation in dense, nonlinear media, a property we will need for the calculation of the transient absorption of light in hybrid nano-materials.

In regions occupied by SPP-sustaining materials (e.g. silver metal) we adopt the conventional Drude model for the dielectric constant of the metal,

$$\varepsilon(\omega) = \varepsilon_r - \frac{\Omega_p^2}{\omega^2 - i\Gamma\omega}, \qquad (5)$$

where $\varepsilon_r$ is the limit of the dielectric constant at high frequencies, $\Omega_p$ is the bulk plasma frequency, and $\Gamma$ is the phenomenological damping rate for this specific metal. For the case being considered here, silver over a frequency range in and near the visible, we use values derived from the work of Gray et al. [33]: $\varepsilon_r = 8.926$, $\Omega_p = 1.760 \times 10^{16}$ rad/sec, $\Gamma = 3.084 \times 10^{14}$ sec$^{-1}$.

The current density $\vec{J}$, which replaces the polarization current $\partial \vec{P}/\partial t$ in Ampere's law in (1), is evaluated according to[34]

$$\frac{\partial \vec{J}}{\partial t} = a\vec{J} + b\vec{E}, \tag{6}$$

where $a = -\Gamma$ and $b = \varepsilon_0 \Omega_p^2$.

The resulting system of partial differential equations is discretized in space and time using the finite-difference time-domain method[34]. Maxwell's equations, along with the Liouville-von Neumann equation, are propagated self-consistently in time at every grid point driven by the local electric field, using the weakly coupled method outlined in Refs. [35]. For all calculations, irrespective of the dimensionality, numerical convergence is achieved at the spatial resolution of $\delta x = 1$ nm and a time step of $\delta t = \delta x/(2c) = 1.7 \times 10^{-3}$ fs, where $c$ is the speed of light in vacuum. To simulate semi-infinite spatial systems we implement the convolutional perfectly matched layers (CPML) absorbing boundaries[36].

Since our main interest is in the optical response of materials much smaller than the incident wavelength, we use incident plane waves. To ensure proper excitation we implement the total-field/scattered-field approach[34] with the following time dependence,

$$E_{inc} = E_0 f(t) \cos(\omega_{inc} t), \tag{7}$$

where the time envelope has the form $f(t) = \sin^2\left(\pi \frac{t}{\tau}\right)$, and $\tau$ is the incident pulse duration. For the probing of transient effects, one typically uses 'white light' with a flat spectrum over the spectral region of interest. We simulate a white light probe with a 0.36 fs long pulse, which produces an essentially flat spectrum for relevant energies between 1 and 4 eV. A probe amplitude of 1 V/m is used throughout this manuscript so as to lie in the weak field limit.

The key challenge in modeling nonlinear dynamics using a pump-probe pulse sequence is to disentangle signals caused by the strong pump from observations by the weak probe. When a system comprised of optically coupled emitters is excited by a strong pulse, it exhibits polarization oscillations lasting long after the pump is gone. Consequently, when the system is probed by a low intensity pulse, one observes an undesired high intensity signal at the pump frequency caused by induced polarization oscillations. Experimentally, such oscillations may be filtered out so as not to interfere with the probe in the far field. In simulations, these unwanted oscillations must be handled carefully because they may interfere with the signal produced by the probe.

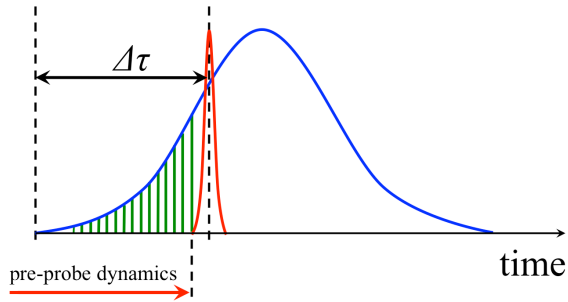

FIG. 1. A high intensity pump (blue) interacts with the system under consideration (see text). At time $\Delta\tau$ after the start of the pump, a low intensity short probe (red) is applied, and its transmission (or absorption) is monitored

Here we propose an efficient computational method to simulate transient spectroscopy experiments. The idea is illustrated in Fig. 1. A high intensity pump drives a system under consideration during some time interval that we denote by $\Delta\tau$, at which time a probe pulse is applied. The Maxwell-Liouville-von Neumann equations are propagated in time and space with the driving pump up until the end of the interval $\Delta\tau$, at which time the density matrix elements are recorded at all grid points where the quantum medium is located. These data are used as initial conditions for simulation of the probe interaction with the sample. This method guarantees that undesired high amplitude oscillations are not included when one probes the system and that the probe does not alter the optical response of the system.

### III. TRANSIENT SPECTROSCOPY OF MOLECULAR NANO-LAYERS

We consider first a thin layer of interacting molecules, depicted schematically in the left inset of Fig. 2. The layer is infinite in the $x$ and $y$ dimensions and finite in $z$. Each molecule in the layer is treated as a two-level system. The layer is subject to external plane wave excitation at normal incidence. The symmetry of the problem reduces the resulting system of coupled equations to the familiar one-dimensional Maxwell-Bloch equations[37]. With the assumption that all molecules are initially in the ground state, the one-photon absorption exhibits a broad resonance near the molecular transition, as shown in the main panel of Fig. 2. This figure displays three absorption spectra, corresponding to different density regimes. One is calculated at a low molecular concentration (blue circles), showing a peak at the molecular transition frequency of 1.61 eV (the "non-interacting" molecules limit); a second corresponds to an intermediate concentration (green squares) and the third is a red-shifted (lower energy) spectrum for a high density sample (red diamonds), where the shift is caused by back action of the molecules on the field, as given by (4). The higher the molecular concentration, the greater the red shift, as expected.

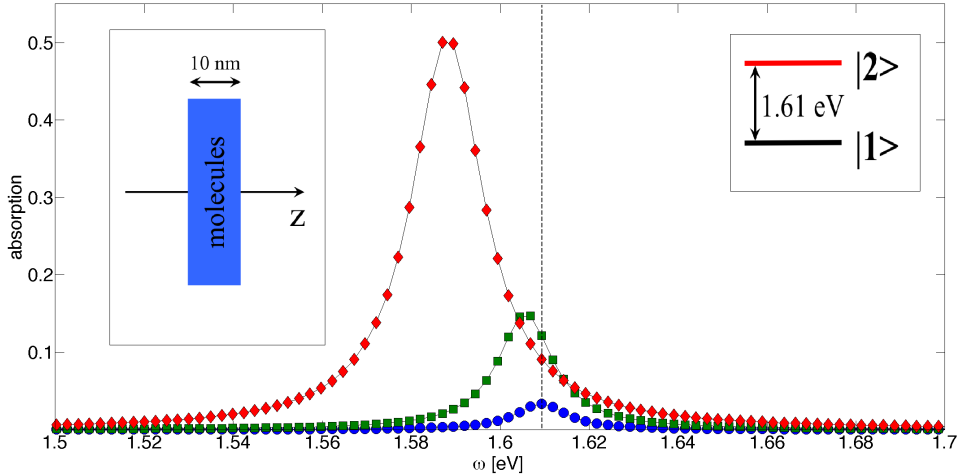

FIG. 2. One-photon absorption for the 10 nm thick molecular layer depicted in the left inset as a function of the incident photon energy calculated at three molecular number densities: blue circles corresponding to $10^{25}$ m$^{-3}$, green squares for $5\times10^{25}$ m$^{-3}$, and red diamonds for $2.5\times10^{26}$ m$^{-3}$. The molecules are modeled as two-level emitters (right inset). The vertical dashed line shows the molecular transition energy of 1.61 eV. In these simulations the molecular transition dipole is 10 Debye, the radiationless lifetime of the excited state |2> is 1 ps, and the pure dephasing time is 100 fs.

We next examine the nonlinear dynamics of such a system under strong pump excitation. As an illustrative example we consider a sequence of $n$ $\pi$ pulses interacting with the molecular nano-layer (with no plasmonic substrate) under the assumption that all molecules are initially in the ground state. Following the numerical procedure discussed in the previous section, we calculate the instantaneous absorption of a probe pulse delayed relative to the pump excitation. Figure 3 shows the results for four different pump pulses, with $n$ increasing from 1 to 4. As anticipated, the molecules undergo Rabi oscillations that depend on the pump pulse area. The absorption becomes negative (indicating gain) when the population is inverted. We also note three important observations: (1) it is more efficient to pump the system at the molecular transition energy, 1.61 eV, than at its red-shifted value of 1.59 eV (confirmed in a set of separate calculations not shown here); (2) the system undergoes transitions back to the ground state with the absorption centered at the red-shifted frequency because of strong mutual interaction between molecules at high concentrations; (3) the Rabi cycling is not complete, and each subsequent oscillation of the system from the ground state to the excited state is less pronounced, an effect that is attributed to decoherence.

Another important factor also contributes to the effect of incomplete Rabi cycling. Even though the thickness of the layer (10 nm) is much smaller than the incident wavelength (770 nm at 1.61 eV), the high molecular density causes the electric field inside the layer to be inhomogeneous; i.e. the local EM field decreases as one probes further within the molecular layer. This decrease results in a lower efficiency of the 'nominal' $n\pi$-pulses. We note that this effect plays a significant role in hybrid systems, as we show in the next section.

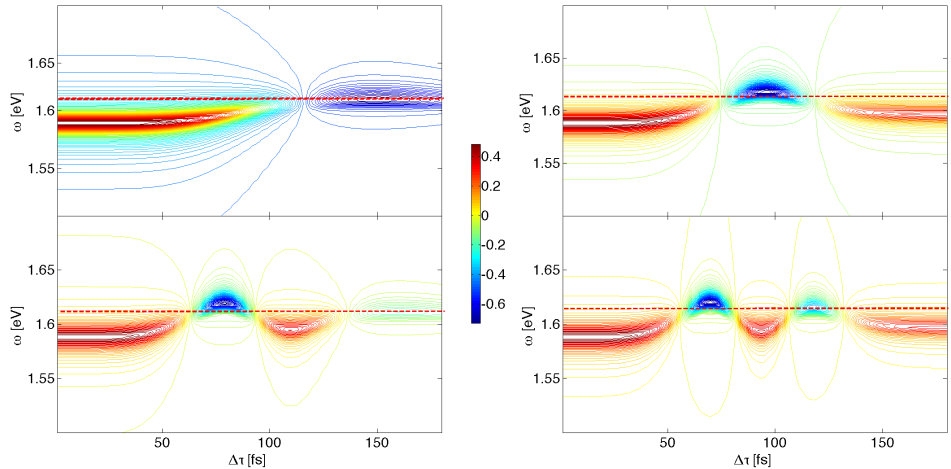

FIG. 3. Transient absorption spectra of the pure molecular system as a function of the pump-probe delay (horizontal axis) and the incident photon energy (vertical axis) evaluated for 180 fs pump pulses. The horizontal dashed black line in each panel is at the molecular transition energy, 1.61 eV. Panel (a) shows the spectrum for a π-pulse ($E_0$ = $2.075 \times 10^8$ V/m), panel (b) shows the data for a 2π-pulse ($E_0$ = $4.150 \times 10^8$ V/m), panel (c) shows results for a 3π-pulse ($E_0$ = $6.225 \times 10^8$ V/m), and panel (d) is for a 4π-pulse ($E_0$ = $8.300 \times 10^8$ V/m). In all cases the pump frequency is resonant with the molecular transition energy. Other parameters are the same as in Fig. 2. A pump-probe delay of 0 fs corresponds to the probe applied at a time when the pump is still off (i.e. at the beginning of the leading edge of the pump; see Fig. 1 for details).

## IV. TRANSIENT SPECTROSCOPY OF PERIODIC HYBRID MATERIALS

The main goals of this paper are to examine the nonlinear optical dynamics in hybrid materials and to probe the influence of surface plasmons polaritons. Motivated in part by recent experimental work[25, 28], we consider a thin molecular layer deposited on top of a periodic array of slits in a silver film, as depicted in the inset of Fig. 4a. In order to account for all possible polarizations of the EM field in the near-field zone[35a], individual molecules are treated as two-level emitters with a doubly degenerate excited state (see the inset of Fig. 4b). The main panels of Figs. 4a and 4b show the linear optical response of the hybrid system at two molecular concentrations. As in the case of a stand-alone molecular layer, we perform white-light transient spectroscopy simulations for normal incidence to compute the transmission coefficient, $T$, and the reflection coefficient, $R$. Since our simulations are performed in the weak probe (linear) regime, one may also calculate the absorption using energy conservation,

$$A = 1 - T - R. \quad (8)$$

Figure 4a shows all three coefficients at low molecular concentration calculated before the pump arrives ($\Delta \tau = 0$). One can clearly see the Rabi splitting in both transmission and reflection. Maxima in transmission correspond precisely to minima in reflection, indicating the energies of the hybrid system, i.e. the upper and lower polaritons. Absorption, on the other hand, has a single narrow resonance with a full width at half maximum corresponding to the energy of the Rabi splitting. (For the given set of parameters, the Rabi splitting is 75 meV). It should be noted that the observed splitting in both transmission and reflection coefficients is indeed the Rabi splitting and is not caused by simple molecular absorption. One may verify this explanation by calculating the effect of varying either the angle of incidence or the periodicity of the slit array on the energies of the upper and lower polaritons[30]. These calculations show strong dispersion along with an avoided crossing behavior – a clear indication of strong coupling between molecular excitons and plasmons.

Figure 4b presents results of simulations at higher molecular density, clearly showing a collective molecular-plasmon mode[30] at 1.61 eV. This energy corresponds to a maximum in $T$ and a minimum in $R$. Note that absorption now has two peaks, which match the positions of the minima of $T$ and maxima of $R$.

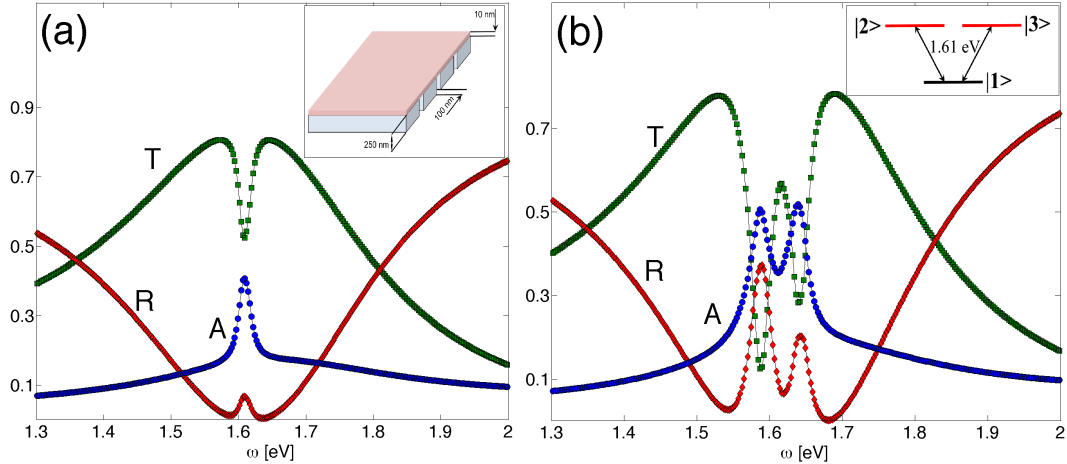

FIG. 4. Extinction spectra of the hybrid material depicted in the inset of panel (a). Both panels show absorption, $A$, (blue circles), transmission, $T$, (green squares), and reflection, $R$, (red diamonds) as functions of the incident photon energy. Individual molecules are considered as two-level emitters with a doubly degenerate excited state, as shown in the inset of panel (b). Panel (a) shows spectra at a low molecular number density of $3\times 10^{25}$ m$^{-3}$. Panel (b) shows the results of simulations at a higher density of $2.5\times 10^{26}$ m$^{-3}$. The slit array period is 410 nm. The other parameters are as in Fig. 2.

Anticipating comparison with experiments, we monitor the nonlinear changes in the reflection spectra rather than the transmission. These changes may be expressed as the ratio,

$$\Delta R(\Delta\tau,\omega) = \frac{R(\Delta\tau,\omega) - R(0,\omega)}{R(0,\omega)}, \qquad (9)$$

where $R(0,\omega)$ is the unperturbed reflection in the absence of the pump.

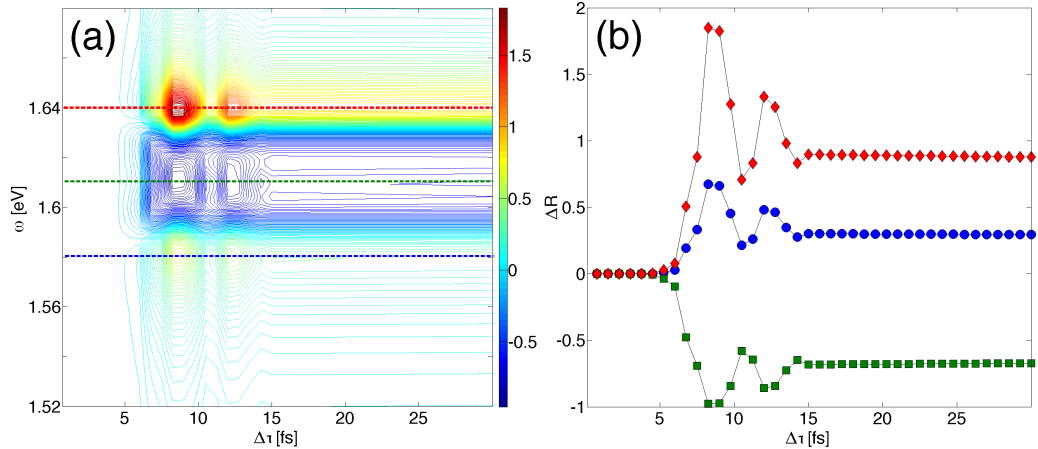

FIG. 5 (Color online) Transient spectroscopy calculations for a 15 fs pump pulse with a peak amplitude of $4\times 10^9$ V/m centered at the molecular transition energy of 1.61 eV. Panel (a) shows the change in reflection $\Delta R$ (see Eq. (9)) as a function of the pump-probe delay and incident photon energy. Panel (b) depicts one-dimensional cuts of $\Delta R$ at three

energies: blue circles are for the lower polariton at 1.58 eV (this energy is also indicated in panel (a) as a horizontal blue dashed line), red diamonds are for the upper polariton at 1.64 eV (also shown in panel (a) as a horizontal red dashed line), and green squares are for the molecular line at 1.61 eV (shown in panel (a) as a horizontal green dashed line). The molecular number density is $3\times10^{25}$ m$^{-3}$ (see Fig. 4a), and the period of the slit array is 410 nm. The other parameters are as in Fig. 2.

Fig. 5a shows results for a 15 fs pump resonant with the molecular transition frequency at 1.61 eV. The observed transient spectra exhibit pronounced oscillations centered about three distinct energies, two of which correspond to the upper and lower polaritons (the minima of the unperturbed reflection, see Fig. 4), while the third is at the molecular transition energy. This result is consistent with recent experimental observations[28], namely, the reflection is increased during the pump at the energies of the upper and lower polaritons. One should also note the asymmetry in the reflection signal – the upper polariton is significantly more enhanced than the lower one. This effect is most likely due to the fact that the unperturbed reflection (Fig. 4a) is already asymmetric, with the molecular transition energy slightly offset towards the upper polariton. Another observation worth noting is that reflection is suppressed at 1.61 eV ($\Delta R < 0$), whereas transmission is enhanced. The molecules that mostly affect both transmission and reflection are located in a close proximity to the slits. Resonant excitation of these spatial regions results in strong coupling of molecular excitons and SPP modes, thereby suppressing reflection and enhancing transmission – a phenomenon similar to extraordinary optical transmission.

To demonstrate that the observed changes are caused by energy transfer between the upper/lower polaritons and the molecular layer, we plot $\Delta R$ in Fig. 5b at three energies as a function of the pump-probe delay. The period of oscillations is 3.75 fs. Interestingly, simulations performed at higher molecular concentrations or different array spacings show that these factors have a very minimal effect on the period of oscillations, which is fully determined by the local electric field amplitude and the strength of the molecular dipole. We also performed calculations using off-resonant pump excitation, as in Ref. [28]. The data obtained for the off-resonant case shows the same behavior – ultra-fast energy oscillations between the upper/lower polaritons and the molecules.

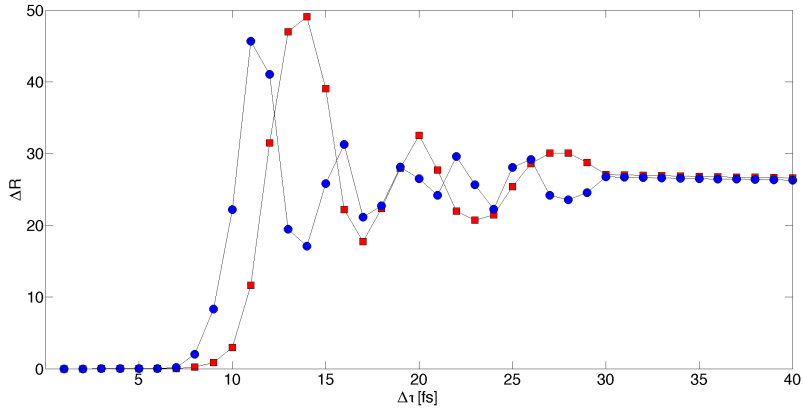

FIG. 6 (Color online) The change in reflection $\Delta R$ as a function of the pump-probe delay at the upper polariton energy for two pump amplitudes: red squares are for $2\times10^9$ V/m (peak amplitude) showing an oscillation period of 6 fs, and blue circles are for $4\times10^9$ V/m with a period of 3 fs. The pump is resonant with the molecular transition energy (1.61 eV), and its duration is 30 fs. The other parameters are as in the previous figures.

Next we proceed to examine the influence of the pump peak amplitude on the nonlinear dynamics. In order to observe many Rabi oscillations cycles but still maintain a moderate incident peak amplitude, we use a pump pulse with a duration of 30 fs. Figure 6 presents $\Delta R$ at two peak amplitudes at the energy of the upper polariton as a function of the time delay. As expected, the Rabi oscillation period decreases with the increase of the pump amplitude. The average period of the oscillations at $4\times10^9$ V/m is ~3 fs, whereas the calculations for $2\times10^9$ V/m indicate a period of ~ 6 fs. These calculations display at least two surprising features. First, the Rabi period is not constant, but rather decreases with time (detailed data not shown). (We note that this variation was also observed in Ref. [28].) Second, the Rabi period obtained from the simulations (the average oscillation period at $2\times10^9$ V/m is 6.9 fs) is smaller than the one obtained assuming a simple two-level atom in the same laser field (17 fs at $2\times10^9$ V/m)[37].

It is incorrect to use the peak Rabi frequency to estimate the oscillation period of the population for a two-level emitter exposed to short pulse excitation. Rather, one has to take carefully into account the laser pulse envelope, calculating the pump pulse area as

$$S = \frac{d_0 E_0}{\hbar} \int f(t) dt, \quad (10)$$

where $d_0$ is the transition dipole moment of the quantum emitter. For observations after the end of the pulse, the integral should be over the entire pulse envelope, whereas for measurements during the pulse, the integral should be taken up to the time of observation by the probe. For an incident source of the form (7), the pulse area becomes

$$S = \frac{d_0 E_0 \tau_{pump}}{2\hbar}. \quad (11)$$

For example, for a 30 fs pump with an amplitude of $2\times10^9$ V/m and the molecular parameters in Fig. 2, we have $S = 1.74\times\pi$. Consequently, a two-level emitter exposed to such an excitation undergoes less than one Rabi cycle. (The population remaining in the ground state is 0.74 after the pump; i.e. the emitter is excited and then partially de-excited). Simple calculations performed for a single two-level emitter agree perfectly with (11).

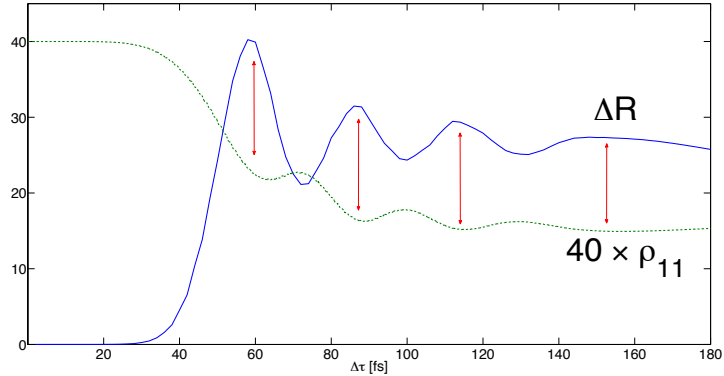

FIG. 7. Nonlinear dynamics during a 180 fs pump. The blue solid curve shows $\Delta R$ as a function of the pump-probe delay at the upper polariton energy. The dashed green curve presents the ensemble-averaged ground state population of the molecular system. (The population $\rho_{11}$ is multiplied by 40 for clarity). Vertical red arrows indicate direct correspondence between oscillations in reflection and the ground state population. The pump amplitude is $4.3\times10^8$ V/m. The incident frequency is at the molecular transition energy of 1.61 eV. The other parameters are as in Fig. 2.

If we now apply this simple model to estimate how many Rabi cycles the hybrid material undergoes, we find that the number of cycles predicted by (11) is always smaller than the actual number computed from the complete set of Maxwell-Liouville-von Neumann equations. The only parameter that is different in the complete model is the local electric field, which differs from its incident value in the near-field zone because of SPPs. The local field enhancement hence plays a crucial role determining how rapidly energy oscillates in hybrid materials.

Simulations using longer pump pulses reveal a direct connection between the excitation dynamics and transient spectra of the system, as illustrated in Fig. 7. There is a clear correspondence between oscillations of the ground state population and changes in reflection induced by the pump. (A similar correlation was found for $\Delta T$.) We conclude, therefore, that the observed oscillations in the transient spectra are due to quantum transitions between ground and excited states in individual molecules.

We conclude this section by examining the role of dephasing in hybrid materials. This property is greatly affected by the inhomogeneous electric field due to SPPs, as is evident from Fig. 7. One of the causes of the fast decay of Rabi oscillations (in addition to pure decoherence and relaxation caused by interaction with the metallic system, which is explicitly included in the model) is spatially dependent excitation of the molecular layer, resulting from the strong gradient of the electric field induced mainly by surface plasmons. Different parts of the molecular ensemble therefore experience different local fields. This effect in turn changes the plasmon dynamics, influencing not only the amplitude of Rabi oscillations but also the Rabi period. It should be noted that the molecular layer influences the near-field via the polarization current, which results in even faster dephasing. Simulations with a thicker molecular layer confirm the aforementioned conclusion, i.e. the Rabi oscillations decay significantly faster when a 10 nm thick molecular layer is replaced by a layer with a thickness of 50 nm.

## V. CONTROL OF THE ENERGY DISTRIBUTION IN HYBRID SYSTEMS

The asymmetry of the energy distribution between the upper and lower polaritons is due to the relative position of the molecular transition energy with respect to the SPP resonance. Such a property of hybrid systems, along with the intriguing experimental possibilities of controlling the structural parameters of plasmonic nanomaterials, call for possible control of the plasmon energy distribution. One may shift the plasmon resonances by changing various material parameters, such as film thickness, period, slit width, etc. Fig. 8 illustrates this idea by examining transient spectra of the hybrid system for different slit periods.

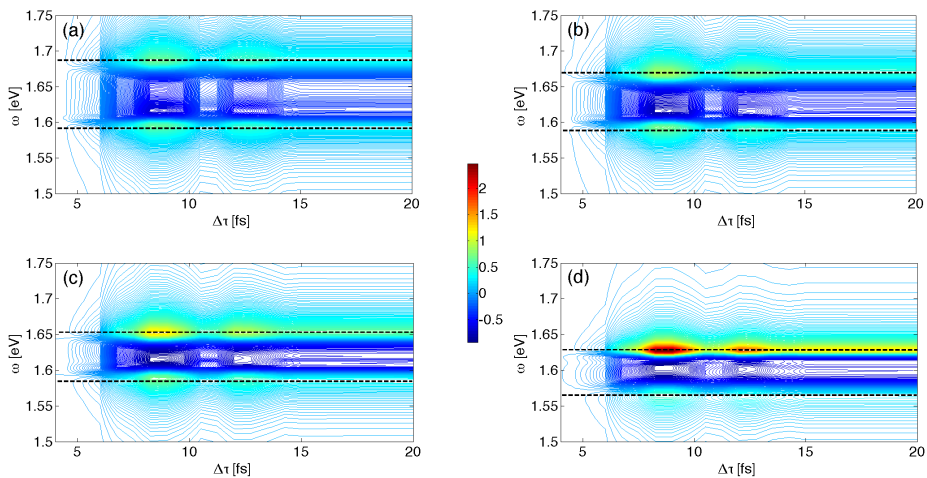

FIG. 8. Change in reflectivity, $\Delta R$, as a function of the pump-probe delay and incident photon energy for different slit array periods: (a) 350 nm; (b) 370 nm; (c) 390 nm; (d) 440 nm. The two horizontal dashed lines in each panel indicate the energy positions of the upper and lower polaritons. The pump pulse duration is 15 fs, the peak amplitude is $4\times10^9$ V/m, the incident frequency is at the molecular line of 1.61 eV, and the molecular number density is $3\times10^{25}$ m$^{-3}$. Other parameters are as in Fig. 2.

The possibility of such control is due to the strong dispersion of the upper and lower polaritons in hybrid systems[19c]. As the plasmon resonance sweeps through the molecular line (for varying slit periodicity), the mixed plasmon-molecular hybrid modes drastically change their positions. The energy is more equally distributed among the upper and lower polaritons at shorter periods, as evident from Fig. 8. One may also envision control by changing the angle of incidence, as this parameter changes the in-plane wave vector, which in turn would change the energy distribution in a fashion similar to varying the array period[38].

A more intriguing control possibility is to utilize the full machinery of the incident laser radiation. To demonstrate how the characteristics of the laser pulse affect the polariton energy distribution, we performed a series of pump-probe simulations varying the time envelop of the pump by turning it off 'suddenly', namely on a time-scale short compared to the natural system time-scales. Figure 9a depicts schematically a control pump with time duration $t_{cutoff}$. Figures 9b-9d show the results of simulations for an array of slits with a period of 370 nm. By controlling $t_{cutoff}$ one may control the number and phase of Rabi cycles, and, as a result, the plasmon energy distribution as well. We can envision simultaneously varying both the time envelope of the pump and its incident angle. The former would control which mode the energy of the system goes to, while the latter manipulates the distribution of energy between the upper and lower polaritons.

An interesting observation is "Free Induction Decay" type oscillations at times $t > t_{cutoff}$, a result of the coherent superposition of the polariton population prepared by the pump pulse before its abrupt termination at $t_{cutoff}$, as seen in Fig. 9. Note that for very short times after the cutoff, the dynamics of $\Delta R$ barely differs from the corresponding dynamics with the pulse on.

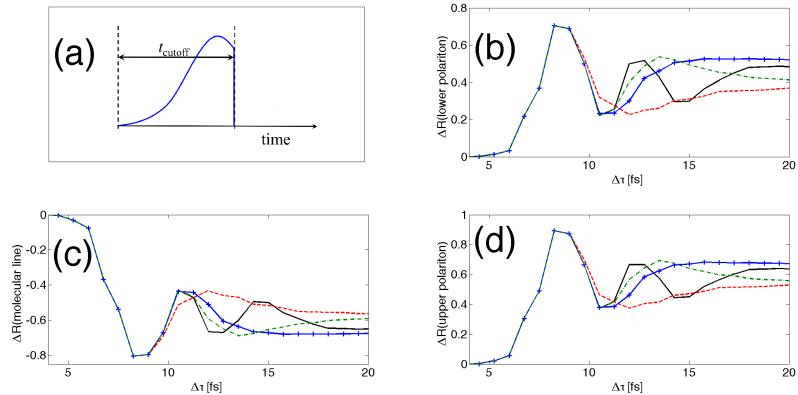

FIG. 9. Effect of the laser pulse shape on the change in reflectivity. Panel (a) illustrates the abrupt cutoff of the pump pulse and application of the probe pulse at time $t_{cutoff}$, which is used as a control parameter. Panels (b)-(d) show $\Delta R$ as a function of the pump-probe delay at different photon energies. Black solid curves show results for a non-truncated pump, red dashed curves show data for $t_{cutoff}$ = 6.75 fs, blue solid curves with crosses are for $t_{cutoff}$ = 7.75 fs, and green dash-dotted lines are for $t_{cutoff}$ = 8.75 fs. Panel (b) shows $\Delta R$ at the lower polariton energy of 1.59 eV, panel (c) shows $\Delta R$ at the molecular transition energy of 1.61 eV, and panel (d) presents $\Delta R$ at the upper polariton energy of 1.67 eV. Simulations are performed for a slit array with a period of 370 nm. The total pump pulse duration is 15 fs, the peak

amplitude is $4\times10^9$ V/m, the incident frequency is at the molecular line of 1.61 eV, and the molecular number density is $3\times10^{25}$ m$^{-3}$. Other parameters are as in Fig. 2.

## VI. CONCLUSION

Using a self-consistent model of the coupled Maxwell-Liouville-von Neumann equations, we scrutinized the nonlinear dynamics of nano-materials comprised of interacting quantum emitters and plasmons. We showed that ultra-short resonant laser pulses significantly modify the optical properties of such hybrid systems. It was demonstrated that energy transfer between the molecular layer and surface plasmons occurs on a femtosecond time scale. This energy transfer may be controlled by altering the material and/or laser parameters. When an intense resonant laser pulse excites a quantum medium coupled to a plasmonic material, the induced spatial distribution of the population of the excited quantum states depends strongly on the incident wavelength and the geometry of the plasmonic material, among other optical and material parameters. If the peak amplitude of the incident field is sufficiently high, the quantum emitters may be driven through one or more Rabi cycles, so that at the end of the pulse they will be in a predetermined quantum superposition of molecular states. The final superposition depends on the local EM field, which is subject to external control (e.g., through variation of the distance from the plasmonic material).

Within the lifetime of the excited states, which may be relatively long for specific systems, the quantum system is being modified by the laser pulse so that its macroscopic refractive index is changed. It should be emphasized that this modification is spatially dependent, with a characteristic length scale much smaller than the incident wavelength. We note that the close proximity of plasmonic materials modifies the refractive index in a spatially dependent manner due to the strongly inhomogeneous near-fields, leading to a modified, highly anisotropic refractive index. One may then probe the system with a low intensity pulse to measure the new refractive index.

Our results suggest a wide variety of future research opportunities, ranging from control of the competition between charge transport and energy transfer – a hurdle in control of light-triggered molecular conduction junctions, to modification of the relationship between light enhancement and excited state quenching – the main handicap of plasmon enhanced spectroscopies.

**Acknowledgements**

M.S. is grateful to the Air Force Office of Scientific Research that partially supported this research via Summer Faculty Research Fellowship 2013. T.S. thanks the NSF (grant No. CHE-1012207/001) for support. RJG thanks the NSF for support under grant No. CHE-0848198.Y.P. acknowledges support by the Israel Science Foundation.